\documentclass[11pt]{article}
 \usepackage{amsfonts}
\usepackage{amsmath} 
 \usepackage{amssymb}
 \newfont{\bbbold}{msbm10 scaled \magstep1}

 \def\cD{{\cal D}}
 
 \def\cF{{\cal F}}

 \newfont{\goth}{eufm10 scaled \magstep1}

 \def\a{\alpha}
 \def\b{\beta}
 \def\c{\gamma} 
 \def\d{\delta}
 \def\e{\epsilon}

 \def\k{\kappa}
 \def\l{\lambda}\def\L{\Lambda}

 \def\p{\pi}

 \def\t{\tau}
 \def\th{\theta}

 \def\ua{\underline{\alpha}}

 \def\una{\underline a}\def\unA{\underline A}

 \def\unM{\underline M}

 \def\be{\begin{equation}}\def\ee{\end{equation}}
 \def\bea{\begin{eqnarray}}\def\eea{\end{eqnarray}}
 \def\ba{\begin{array}}\def\ea{\end{array}}
 
\newcommand{\cc}{\gamma^{\left(2\right)}}
\begin{document}
 %%%%%%%%%%%%%%%%%%%%%%%%%%%%%%%%%%%%%%%%%%%%%%%%%%%%%%%%%%%%%%%%%%%%%%%%%%%%%
\bibliographystyle{utcaps}
 \thispagestyle{empty}

 \hfill{KCL-TH-02-09}

 \hfill{hep-th/0204225}

 \hfill{\today}

 \vspace{20pt}

 \begin{center}
 {\Large{\bf Supersymmetric Born-Infeld from the D9-brane}}
 \vspace{30pt}

 {Sven F. Kerstan}
\vskip 1cm {Department of Mathematics} \vskip 1cm {King's College,
London} \vspace{15pt}

\end{center}

 \vspace{60pt}

 {\bf Abstract}
Using the superembedding approach, the full supersymmetric effective field theory of the D9-brane, super Born-Infeld theory,
is fixed by the so called $\cF$-constraint. The odd-odd components of the theory's super field strength, $f_{\a \b}$, are implied by this 
constraint. Given $f_{\a \b}$, the super Bianchi identities imply the theory's equations of motion.\\
We calculate $f_{\a \b}$ up to order 5 in fields, corresponding to order 6 in fields in the Lagrangian.

{\vfill\leftline{}\vfill \vskip  10pt

 \baselineskip=15pt \pagebreak \setcounter{page}{1}

 %%%%%%%%%%%%%%%%%%%%%%%% INTRODUCTION %%%%%%%%%%%%%%%%%%%%%%%%%%%%%%%%%%%%%%
 %%%%%
\section{Introduction}

In the 1930s, Einstein's mechanism to introduce a maximal velocity in special relativity, the $\sqrt{1-v^2}$ term, motivated Born and 
Infeld to try and introduce a maximal field strength in Maxwell's theory in the same way. The result was what we now call
Born-Infeld (BI) theory.\\
In the 80s, the BI action was shown to arise in open bosonic string theory \cite{fradkintseytlin}. After the
discovery of D-branes, it was realised that BI theory is a key ingredient in the effective theory of branes \cite{leigh}. 
Remarkably, this enables us to understand BI theory as not merely motivated by special relativity, but in fact as T-dual to it 
\cite{bachas}:
a relativistic point particle is the T-dual description of a brane with a slowly varying electric field E.
In a recent work \cite{defossekoerbersevrin}, it was shown that demanding the existence of stable holomorphic vector bundles 
(corresponding to instantons in Yang-Mills theory) to be solutions of a deformed abelian Yang-Mills theory fixes deformations
to be BI theory. Another property of BI theory is that it is unique in its form invariance under the Seiberg-Witten map \cite{wyllard}.\\
Supersymmetrisation of BI theory began in the later 1980s, when in \cite{cecottiferrara} a N=1 supersymmetric BI Lagrangian in D=4
was presented. Later, manifest extended supersymmetry was considered (e.g. N=2, D=4 in \cite{ketovabelian}). In the later 90s,
partially broken global supersymmetries (PBGS) were investigated. For example, in \cite{baggergalperin} it was shown that 
partially breaking N=2 in D=4 can be solved by the Maxwell multiplet, resulting in BI theory, while the breaking of N=2 in
D=10 was considered in \cite{bellucciivanovkrivonos}.
A geometric approach to PBGS is the superembedding method, in which one superspace (interpreted as brane world volume) 
is embedded into another, larger one (target space), using appropriate constraints ( the so called embedding condition and
the $\cF$-constraint). The result is again BI theory. The virtue of the superembedding approach is not only that it gives
a geometrical interpretation to the theory and illuminates the nature of $\kappa$-symmetry 
\cite{volkovzheltukhin,stv,stvz}, but also that it leads to a systematic way of determining the effective action for branes in
various dimensions \footnote{For an earlier attempt to use source and target superspaces, see, e.g. \cite{gates1}.} 
\cite{bandossorokinvolkov,howeraetzelsezgin}. For a review of the superembedding method see \cite{sorokin}. For examples of
derivations of D-brane dynamics from superembeddings, see \cite{howesezgin,howeraetzelsezgin,bppst,drummondhowe}.
The equivalence between PBGS and the superembedding approach (for flat target superspace) has been discussed in \cite{pst,pst2} and,
in particular has been shown for space-filling branes in 3 and 4 dimensions in \cite{bppst,drummondhowe}.
More recently it has been shown (for D=3 and D=4) that the constraint used in the nonlinear realisation approach
\cite{ivanovkrivonos,baggergalperin} is equivalent to the 
$\cF$-constraint in the superembedding approach and that the resulting actions are identical \cite{james}.\\
The search for a non-abelian version of super BI (NBI) theory, which would correspond to a theory describing coinciding branes, is
an active area of current research. The NBI action had been investigated at order $\alpha^{\prime 2}$ in the mid eighties
\cite{grosswitten,yetanothertseytlin,bergshoeffrakowskisezgin}. Later, symmetrised trace actions were proposed
\cite{tseytlin,ketovnonabelian}. Recent calculations suggest that these proposals are not valid beyond order 4 in $F$
\cite{hashimototaylor}. Other approaches taken towards the higher order terms of NBI are considering non-abelian $\kappa$-symmetry 
\cite{bergshoeffderoosevrin} which was used to derive a bosonic and fermionic terms up to order 4 in $F$, calculations 
of correlators for D=4, N=4 quantum Yang-Mills (up to order 5 in $F$) \cite{zanon}, and the deformation of Yang-Mills, given that stable 
holomorphic bundles still are solutions to the deformed theory \cite{koerbersevrinNA,koerbersevrinNA2}.
A calculation from string scattering amplitudes which gave the action up to order four in fields can be found in \cite{bergshoeff}.
Apart from these approaches, there have been attempts to work out in how far supersymmetry fixes possible NBI theories \cite{cnt1,cnt2,cnt3}.\\

In this paper we will make use of the (flat) superembedding formalism to determine the superspace form of the effective field theory
for an abelian D9-brane. Since the multiplet is on shell and there is no known way to force the theory off-shell, we do not 
write down a Lagrangian for this theory. Instead, we calculate the odd-odd part of the super field strength, $f_{\a \b}$, which is implied
by the so called $\cF$-constraint. We will write $F$ for the world volume field strength 2-form, $F_{AB}$ for its components in the 
basis induced by our embedding, and $f_{AB}$ for its components in a flat basis. So $f_{AB}$ is the normal N=1, D=10 field strength 
satisfying $df=0$  in flat superspace. The constraint on $f_{AB}$ which produces super Maxwell theory is $f_{\a \b}=0$. 
The lowest order deformation (order three in fields) of this constraint has been known for a long time 
(\cite{gates2,gates3,bergshoeffrakowskisezgin}) and in conventional fields ($f_{ab}$ the spinor $\lambda$) on dimensional grounds has to be 
$$\left(\c^{edcba}\right)_{\a \b} \left(\l\c_{abc}\l\right) f_{de}$$ 
In \cite{cnt1} it is shown that linear supersymmetry leads to this result as well. Once $f_{\a \b}$ is known, the Bianchi identities 
for the field strength determine the equations of motion.

Our approach gives a closed expression for $f_{\a \b}$ in  fields as they arise in the embedding approach. Redefining these, we 
expand the answer in conventional fields up to order 5.
For consistency checks of proposals of non abelian versions of BI, one can directly compare the abelian limit of $f_{\a \b}$, or, in
case $f_{\a \b}$ is not given (e.g. \cite{bergshoeff}), the resulting equations of motion.\\
Work on the D9-brane has been published before. In \cite{aganagicpopescuschwarz}, the Wess-Zumino term required to 
achieve $\kappa$-symmetry for Dp-brane actions was calculated. In \cite{bandits}, the D9-brane was analysed in a superembedding
context, but the analysis was performed in a linearised approximation.\\
We shall consider the embedding $s$ of a D=10, N=1 world volume with a gauge field super two form $F=dA$ into a flat D=10, N=2 target 
space with a three form super field strength $H=dB$. Treating the world volume as a D-brane, we know from string theory 
that $\k$-symmetry has to be preserved, which, as was pointed out in \cite{Fconstraint}, implies
\begin{itemize}
\item{the embedding constraint: $E_\a{}^{\underline{a}}=0$}
\item{the $\cF$-constraint: $\cF_{\a B} := F_{\a B} - s^*\underline{B}_{\a B} = 0$}
\end{itemize}
where $s^*\underline{B}$ is the pullback of $B$ and $\cF = F-s^*B$.
Implementing these will determine the explicit form of $f_{\a \b}$ which we then express in the 
standard fields (in our case, the embedding constraint does not actually constrain $F$). 
As pointed out before ( e.g. \cite{cnt1}), this determines the equations of motion via the Bianchi identities. Since 
the embedding brakes half the supersymmetry there is a Goldstone degree of freedom, which is the location of the brane in the transverse
odd target space. The resulting theory will then realise one manifest linear (all our fields are N=1 super fields) and one non-linear 
supersymmetry. The $\cF$-constraint is the only covariant constraint compatible with the Maxwell multiplet. This means that the theory
we derive, with one linear and one non-linear supersymmetry, is unique.\\
\newline
In section \ref{embrev}, we give a very brief review of the formalism of superembeddings. In the following section we then consider 
the possible flat backgrounds $H$, calculate their potentials $B$, pull $B$ back to the brane, impose
the $\cF$-constraint and so arrive at an expression for $f_{\a\b}$ in embedding fields. In the next section we then expand this expression
up to order 5 in conventional fields. We conclude with some remarks on our result.

\section{The Superembedding formalism} \label{embrev}

A superembedding is a map from the world volume of the brane to the target space: $s: M \rightarrow \underline{M}$. Letters from the 
middle of
the alphabet will be used for coordinate indices, letters from the beginning for tangent space indices. Lower case Latin letters will 
refer to the even, Greek letters to odd components. Upper case Latin letters include even and odd indices. So we have
$M = \left(m, \mu\right)$ for coordinate indices and $A = \left(a,\a\right)$ for tangent space indices. When referring to the target 
space, indices will be underlined. Coordinates are denoted as $z^M=\left(x^m,\th^\mu\right)$. The relation between tangent and 
coordinate bases is given by 
the super vielbein $E_M{}^A$ and its inverse $E_A{}^M$. Coordinates transverse to the world volume will be primed: $A^\prime$. The 
embedding matrix is the derivative of $s$ in terms of tangent indices:
\begin{eqnarray}
E_A{}^{\unA} := E_A{}^M \partial_M z^{\unM} E_{\unM}{}^{\unA}
\end{eqnarray}
As an example we pull back a one form $Q$ from the target space to the world volume:
\begin{eqnarray}
\left(s^*Q\right)_A = E_A{}^{\unA} Q_{\unA}
\end{eqnarray}
The basic embedding condition one imposes is that the (tangent space of the) odd world volume sits in the (tangent space of)
odd target space: 
$E_\a{}^{\underline{a}}=0$. This is a consequence of demanding $\k$-symmetry for open strings, ending on the brane, as was shown
in \cite{Fconstraint}, and is, without loss of generality, true for the space filling brane we will consider later.
We will denote the standard flat basis of superspace by
\begin{eqnarray}
e^\a &=& d\theta^\a \\
e^a &=& dx^a - \frac{i}{2}d\th^\a \left(\c^a\right)_{\a \b} \th^\b
\end{eqnarray}
The $\c$-matrices are 16 by 16 real symmetric. Locally on the brane it is always possible to choose an induced basis defined by
\begin{eqnarray}
E^\a &=& e^\a \label{flatbasis1} \\
E^a &=& \left(e^b - e^\b \psi_\b{}^b\right) \left(B^{-1}\right)_b{}^a \label{flatbasis2}
\end{eqnarray}
where
\begin{eqnarray}
\psi_\a{}^a = \frac{i}{2} D_\a \th^\prime \c^b \th^\prime \left( \d_b{}^a - \frac{i}{2} \partial_b\th^{\prime}\Gamma^a\th^\prime\right)^{-1}
\end{eqnarray}
and
\begin{eqnarray}
B_a{}^b = \left( \d_a{}^b - \frac{i}{2} \partial_a\theta^\prime\c\theta^\prime\right)^{-1}
\end{eqnarray}
The world volume spinor derivative in the induced basis is
\begin{eqnarray}
\cD_\a := D_\a + \psi_\a{}^a \partial_a
\end{eqnarray}

Finally we may split up the target space into two parts of same dimension such that
\begin{eqnarray}
E_\a{}^{\ua} = u_\a{}^{\ua} + h_\a{}^{\b^\prime} u_{\b^\prime}{}^{\ua}
\end{eqnarray}
$u$ then is an element of a group $\underline{G}$, which will either be the Spin group, or a product of Spin
with some internal symmetry group. The even-even part can then be parametrised by the Lorentz transformation
corresponding to the Spin transformation
\begin{eqnarray}
E_a{}^{\una}=u_a{}^{\una}
\end{eqnarray}

A symmetry transformation $g \in \underline{G}$ will map the frame $E_\a$ to itself, up to a linear transformation, if we have:
\begin{eqnarray}
u \rightarrow gu\\
h \rightarrow h^\prime
\end{eqnarray}
for
\begin{eqnarray}
h^\prime_\a{}^{\b^\prime} = \left(-g_\a{}^{\c^\prime}+ g_\a{}^\d h_\d{}^{\c^\prime} \right)
\left(g_{\c^\prime}{}^{\b^\prime} - g_{\c^\prime}{}^\e h_\e{}^{\b^\prime} \right)^{-1} \label{htrans}
\end{eqnarray}
as a straightforward calculation shows. So the field $h$ transforms projectively under $\underline{G}$.\\
We will choose to align the world volume coordinates with the target space coordinates:
\begin{eqnarray}
x^a &=& x^{\underline{a}}\\
\theta^{\a} &=& \theta^{\underline{\a}1}\\
\Lambda^\a\left(x,\theta\right) &=& \theta^{\underline{\a}2}
\end{eqnarray}
This is called static gauge. $\Lambda$ corresponds to the odd transverse location of the brane in target space. Note that
this fixing of gauge kills the world volume diffeomorphism invariance which implies the loss of $\kappa$-symmetry.\\
For a more detailed introduction to superembeddings, see, for example, \cite{sorokin}, and for the special case of codimension 0, see, for
example, \cite{howeraetzelsezgin}.

\section{Background fields} \label{backgroundfields}

We will embed an N=1, D=10 space, our brane's world volume, in an N=2, D=10 target space. On the world volume there lives a super 
gauge field strength $F$. The embedding condition $E_\a{}^{\underline{a}}=0$ does not constrain $F$, so to fix the multiplet and
the dynamics, we need some other constraint. Demanding $\k$-symmetry for open strings ending on our world volume implies such a
constraint for $F$, the $\cF$-constraint. To implement the $\cF$-constraint, we define a new super fieldstrength $\cF$ on the brane:
\begin{eqnarray}
\cF = F - s^*\underline{B} \label{cFdef}
\end{eqnarray}
and then impose the $\cF$-constraint
\begin{eqnarray}
\cF_{\a \b} &=& 0\\
\cF_{\a b} &=& 0
\end{eqnarray}
With $H$ being the field strength of the super two form field $B$, the Bianchi identity for $\cF$ is:
\begin{eqnarray}
d\cF = -H
\end{eqnarray}
Rearranging (\ref{cFdef}) gives
\begin{eqnarray}
F = \cF + s^*\underline{B} \label{whatsF}
\end{eqnarray}
which will, once we know $s^*\underline{B}$, fix the odd-odd and odd-even components of $F$, due to the $\cF$-constraint. The equations
of motion are then implied by the Bianchi identity $d\cF = -H$.\\
In this section we will now proceed as follows:
\begin{itemize}
\item{determine possible $\underline{H}$-backgrounds}
\item{calculate the corresponding gauge potential $\underline{B}$}
\item{pull $\underline{B}$ back to the brane in a flat basis}
\item{use (\ref{whatsF}) to determine the odd-odd and even-odd components of $F$ (section \ref{getf})}
\end{itemize}
What then remains to be done is to expand the result in the conventional fields, which we do in penultimate section.

\subsection*{Determining $H$} \label{H}
Working with a flat $H$ background, $H$ has to be built from super invariants, i.e. gamma matrices only:
\begin{eqnarray}
\underline{H} = \left( -i \right) E^{\c_j} \wedge E^{\b_i} \wedge E^a \left(\c_a\right)_{\b \c} M_{ij}
\end{eqnarray}
where $M_{ij}$ is some symmetric 2x2 matrix. A basis for these is given by the two symmetric Pauli matrices
and the identity. The Bianchi-identity
\begin{eqnarray}
0 = d\underline{H} 
\end{eqnarray}
gives a further restriction. In components it reads
\begin{eqnarray}
0 = \left(\c_a\right)_{\left( \a_i \b_j \right.} \left(\c^a\right)_{\left. \c_k \delta_k \right)}
\end{eqnarray}
where i and j are restricted by $M_{ij}$ as above. Using
\begin{eqnarray}
\left(\c^{}_a\right)_{\a \left( \b \right.} \left(\c^a\right)_{ \left. \c \delta \right) } = 0 \label{thesym}
\end{eqnarray}
one finds that $M$ cannot be the identity. Further, one finds that both, $\t_1$ and $\t_3$ are possible choices for $M$.

\subsection*{Determining $B$} \label{B}
The aim is to find a two form $B$ on the worldvolume which is only dependent on $\L$. The reason for that is, that
via the $\cF$-constraint, $B$ enters the brane field strength $F$, which contains the physical fields. The only physical
degrees of freedom of the brane are its location in the transverse odd dimension, $\L$, (its $\theta^2$ coordinate in target space), 
and its super partner, the bosonic gauge field on the brane. So, in a physical gauge, $B$ should be built only from 
these and not depend on the brane coordinates $x$ and $\theta^1$. Note that this is just a choice of gauge.
Since the embedding matrix will not change the $x$ or $\theta^1$ dependences, the target space $B$ must only depend on $\L$.\\
Consider $\underline{H}$ to be of $\t^1$ type first. The equation
\begin{eqnarray}
d\underline{B} = \underline{H}
\end{eqnarray}
gives eight component equations, and these can be solved by
\begin{eqnarray}
\underline{B}_{AB} = \left( \begin{array}{ccc} 
0 & -3 i \L^{\epsilon} \left(\c_a\right)_{\epsilon \b} & 0  \\
3i \L^{\epsilon} \left(\c_b\right)_{\epsilon \a} & 0 &
 \l^{\epsilon}\left(\c^f\right)_{\epsilon \a} \left(\c_f\right)_{\b \delta} \L^{\delta}  \\
0 & \L^{\epsilon}\left(\c^f\right)_{\epsilon \a} \left(\c_f\right)_{\b \delta} \L^{\delta} & 0\\
\end{array} \right)
\end{eqnarray}

Considering $\underline{H}$ in $\tau_3$ gauge, we find from the target space component equation
\begin{eqnarray}
D_{\left( \a \right.} B_{\b c \left. \right)} + T_{\left( \right. \a \b}{}^f B_{fc \left .\right)} = H_{\left( \a \b c \right)}
\end{eqnarray}
that the even-even $B$ would have to be proportional to the unit-matrix which is not antisymmetric. So there are no solutions independent
of $\theta^2$ for the $\t_3$ case. Using the embedding matrix to pull $\underline{B}$ back to the world volume, and switching to a flat
basis using (\ref{flatbasis1}), (\ref{flatbasis2}) we find
\begin{eqnarray}
b_{\a \b} &=& \frac{1}{3} D_{\left( \right.\a} \L \c^b \L \left( \c_b\right)_{\b\left.\right) \c} 
\L^\c \\
b_{\a b} &=& -i \left( \c_b\right)_{\a \c} \L^\c- \frac{1}{6} \partial_b \L \c^c \L \left( \c_c
\right)_{\a \c} \L^\c \\
b_{a b} &=& 0 
\end{eqnarray}
where $b$ denotes the pullback of $\underline{B}$ in a flat basis.

\subsection*{Calculating $F$} \label{getf}

Using (\ref{whatsF}) and the just derived components of $B$, we find for the components of $F$ in a flat basis:
\begin{eqnarray}
f_{\a \b} &=& \frac{1}{3} D_{\left( \right. \a} \L\c^b \L \left(\c_b\right)_{\b\left.\right)\mu}
\L^\mu + \left( \frac{i}{2} D_\a \L \c^c \L \right) \left( \frac{i}{2} D_\b \L \c^d \L \right)
\cF_{cd} \label{foddodd} \\
f_{\a b} &=& -i\c_{b \a \b} \L^\b -\frac{1}{6} \partial_b \L.\c^c.\L
\c_{c \a \b} \L^\b - D_\a \L \c^c \L \left( \delta_b{}^d - \frac{i}{2} \partial_b \L
\c^d \L \right) \cF_{cd} \label{foddeven} \\
f_{ab} &=& \left( \delta_a{}^c - \frac{i}{2} \partial_a \L \c^c \L
\right) \left( \delta_b{}^d - \frac{i}{2} \partial_b \L \c^d \L
\right) \cF_{cd} \label{feveneven}
\end{eqnarray}
The first of these equations is, in closed form, the constraint that determines the full 
super Born-Infeld theory. Note that the conventional constraint,  i.e. $f_{\a \b}$ be gamma traceless, has not yet been imposed on $f$.
\section{Redefining the fields}\label{howto}
After having obtained the constraint on $f_{AB}$  in terms of $\cF_{ab}$, which is in the induced basis, and the spinor field $\Lambda$, we
now want to express it in conventional fields. These are contained in the flat world volume field strength $\hat{f}$
which is $F$ in the flat basis with the conventional constraint imposed:
\begin{eqnarray}
\left(\c^a\right)^{\b \a} \hat{f}_{\a \b} &=& 0 \label{convconst} \\
\left(\c^b\right)^{\a \b} \hat{f}_{b \b}&=& 10 \l^\a\label{convspin}
\end{eqnarray}
To perform the necessary field redefinitions, we will proceed as follows
\begin{itemize}
\item{express spinor derivatives of $\L$ in terms of the embedding field h}
\item{express the embedding field h in terms of $\cF$}
\item{express $\cF$ in the flat basis to get $f_{AB}$}
\item{determine the shift in the gauge potential $A \rightarrow \hat{A}$, such that its fieldstrength $\hat{f}_{AB}$ satisfies 
the conventional constraint (\ref{convconst})}
\item{using (\ref{convspin}), express the spinor field $\L$ in terms $\cF$ and the conventional spinor $\l$}
\item{use these relations in the expression for $f_{\a \b}$}
\item{in the resulting equation, $f_{\a \b}$ will appear on the rhs in a higher order expression. Substitute the lowest order
term of the rhs for it, to eliminate (at this order) all $f_{\a \b}$ on the rhs}
\item{impose the conventional constraint on the equation}
\item{simplify the result}
\end{itemize}
First consider the relation between $D_\a\L^\b$ and $h_\a{}^\b$. We know that
\begin{eqnarray}
\mathcal{D}_\a \L^\b = h_\a{}^\b
\end{eqnarray}
and since $\mathcal{D}_\a = D_\a + \psi_\a{}^c \partial_c$ we know that
\begin{eqnarray}
D_\a \L^\b &=&  \mathcal{D}_\a \L^\b - \psi_\a{}^c \partial_c \L^\b \\
&=& h_\a{}^\b - \frac{i}{2} D_\a \L \c \L \left( \delta - \frac{i}{2} \partial \L \c \L
\right)^{-1} \partial \L\\
&=& h_\a{}^\b - \frac{i}{2} \left( h_\a{}^\b + \frac{1}{4} D_\a \L \c\L \left( \delta - \frac{i}{2} 
\partial \L \c \L \right)^{-1} \partial \L \right) \c\L \left( \delta - \frac{i}{2} \partial 
\L \c \L \right)^{-1} \partial \L
\end{eqnarray}
The recursion will terminate after 16 steps, since in each recursion the term with the spinor derivative gets one additional spinor.
It turns out that, for $f$ up to order 5 in fields, we need to determine $D_\a \L^\b$ up to order three in fields, and up to order
three we find
\begin{eqnarray}
D_\a \L^\b \stackrel{*}{=} h_\a{}^\b - \frac{i}{2} \left(h \c \L \partial \L \right)_\a{}^\b \label{hder}
\end{eqnarray}
where $\stackrel{*}{=}$ indicates that the expression on the rhs is only correct up to order three in fields.
\subsubsection*{Relation between $F$ and $h$}
In \cite{bandits} a relation between $F$ and $h$ was derived. Since $h$ is part of the pullback matrix, one may hope
to find this relation when analysing the odd-odd-even component of the deformed Bianchi identity $d\cF = -H$ on the world volume.
Working in the $\tau_1$-gauge (recall that we chose this gauge since the alternative $\tau_3$-gauge does not allow 
pure $\theta^2$ dependence of our fields), one easily sees that this will not work out. Since we know how $h$ transforms 
(\ref{htrans}), however, we may work in the the alternative $\tau_3$-gauge and then transform the result back into $\tau_1$-gauge.
So first we write down the odd-odd-even part of the pullback of the deformed Bianchi identity in induced basis:
\begin{eqnarray}
h_\a{}^\c h_\b{}^\d \left(\c^a\right)_{\c \d} \cF_{ab} + \left(\c^a\right)_{\a \b} \cF_{ab} = \left(\c_b\right)_{\a \b} 
- h_\a{}^\c h_\b{}^\d \left(\c_b\right)_{\c \d} \label{bandostrick}
\end{eqnarray}
Multiply with $\left(\c^{\left(5 \right)}\right)^{\a \b}$:
\begin{eqnarray}
\left(\c^{\left(5 \right)}\right)^{\a \b} {h_\a}^\c {h_\b}^\d \left(\c^b\right)_{\d \c} \left(\cF_{ba} + \eta_{ba}\right) = 0
\end{eqnarray}
If $\cF+\eta$ is invertible, then this implies
\begin{eqnarray}
\left(\c^{\left(5 \right)}\right)^{\a \b} h_\a{}^\c h_\b{}^\d \left(\c^b\right)_{\c \d} = 0 \label{nog5}
\end{eqnarray}
Since $\c$ and $\c^{\left(5\right)}$ form a basis for the symmetric 16 by 16 matrices, (\ref{nog5}) implies that we can
write $ h h \c$ as:
\begin{eqnarray}
{h_\a}^\c {h_\b}^\d \left(\c^b\right)_{\d \c} = \left(\c^f\right)_{\a \b} {n_f}^b
\end{eqnarray}
This tells us that $h$ is in the spinor representation of the Lorentz group (plus additional scaling), 
and $n$ is in the vector representation of the Lorentz-Group (plus scaling). Plugging the last equation into (\ref{bandostrick}) gives
\begin{eqnarray}
&{}&\c^b_{\a \b} \left( \cF_{ba} + {n_b}^c \cF_{ca} \right) = \c^b_{\a \b}
\left( \eta_{ba} - n_{ba}\right)\\
&\Rightarrow& \eta - \cF = n\left( \eta + \cF \right)\\
&\Rightarrow& n^t n = \eta \\
&\Rightarrow& n \in O(1,9) 
\end{eqnarray}
Moreover
\begin{eqnarray}
det\left( \left( \delta - \cF \right) \left( \delta + \cF \right)^{-1} \right) &=& \frac{det\left( \delta - \cF \right)}
{det\left( \delta + \cF \right)} \\
&=& 1
\end{eqnarray}
so we have
\begin{eqnarray}
n \in SO(1,9)
\end{eqnarray}
Note that this result agrees with that in \cite{bandits}, as in the induced basis we have $B_{ab}=0$, so $\cF_{ab}=F_{ab}$.
Now we may write $n$ in terms of the generators $M$ of the Lie algebra:
\begin{eqnarray}
&{}& n_a{}^b = \exp\left( c_{mn} {M^{mn}} \right)_a{}^b\\
&\Rightarrow& \left(\ln n\right)_a{}^b = c_{mn} \left( M^{mn}\right)_a{}^b \\
&\Rightarrow& c_{rs} = c_{mn} M^{mn} .. M_{rs} = \ln\left(n\right) .. M_{rs}
\end{eqnarray}
where $..$ indicates a suitable inner matrix product, such that the generators $M$ form an orthogonal basis.
For the standard Lorentz generators we are using, the trace over the dot product is a suitable inner product.\\
Since we know that the same transformation is realised in $h$ as a spinor representation, the generators of
which are $ \frac{1}{2}i\cc$, we get in $\t^3$ gauge
\begin{eqnarray}
\hat{h}_\a {}^\b &=& \left( \exp \left( \frac{i}{2} c_{rs} \c^{rs}\right) \right)_\a {}^\b\\ 
&=& \left( \exp \left( \frac{i}{2} \ln\left(n\right)  M_{rs} \c^{rs}\right) \right)_\a {}^\b\\
&=& \left( \exp \left( \left(2 \sum_{i=0}^{\infty} \frac{ \cF^{2i+1}}{2i+1} \right)_{rs} 
\c^{rs} \right) \right)_\a {}^\b
\end{eqnarray}
By suitable Lorentz transformations one can always block-diagonalise $\cF_{ab}$, and therefore we do not have
to worry about the non-commutativity of the $\cc$. \\
Remember that the analysis of the embedding field $h$ started using the background field $H$ in the $\t^3$ 
gauge. As stated in section \ref{H}, we want to work in $\t^1$ gauge, though. The transformation relating $H_{\t^3}$
and $H_{\t^1}$ is a rotation between $\theta^1$ and $\theta^2$ by $\frac{\p}{4}$. We observe that
this rotation is an element of the group $G$ (see section \ref{embrev}), and we therefore can use equation (\ref{htrans}) 
to determine the embedding field $h$ in $\t^1$ gauge:
\begin{eqnarray}
h_\a{}^\b &=& -\left( \left( 1 - \hat{h} \right) \left(1+ \hat{h} \right)^{-1}
\right)_\a{}^\b\\
&\stackrel{*}{=}& \cF\cc - \frac{1}{3} \cF\cF\cF \cc\cc\cc+ \frac{1}{3} \cF^3 \cc
\end{eqnarray}
where $\stackrel{*}{=}$,as before, indicates that terms of order greater than three in fields are truncated on the rhs.

\subsection*{$\cF_{ab}$ in terms of $\hat{f}_{ab}$ }

For the even-even part, the pullback of $\cF$ is rather simple:
\begin{eqnarray}
\cF_{ab} &=& B_a^c B_b^d f_{cd} \\
&=&  \left( \delta_a^c - \frac{i}{2}\partial_a \L \c^c \L \right)^{-1}
\left( \delta_b^d - \frac{i}{2}\partial_b \L \c^d \L \right)^{-1} f_{cd}
\end{eqnarray}
To impose the conventional constraint, i.e. make $f_{\a \b}$ $\c$-traceless, we shift the gauge potential
\begin{eqnarray}
f_{\a \b} &=& D_\a A_\b + D_\b A_\a -i \left( \c^a\right)_{\a \b} A_a\\
A_a &\rightarrow& \hat{A}_a = A_a + \frac{1}{16i} \left( \c_a\right)^{\a \b} f_{\a \b}\\
\Rightarrow \hat{f}_{\a \b} &=& f_{\a \b} - \frac{1}{16} \left(\c^a\right)_{\a \b} \left(
\c_a\right)^{\c \delta} f_{\c \delta}
\end{eqnarray}
and so
\begin{eqnarray}
\hat{f}_{\a b} &=& f_{\a b} + \frac{1}{16 i}D_\a \left( \c_b\right)^{\c \delta} f_{\c \delta}\\
\hat{f}_{ab} &=& f_{ab} + \frac{1}{16i}D_a \left( \c_b\right)^{\c \delta} f_{\c \delta}
- \frac{1}{16i}D_b\left( \c_b\right)^{\c \delta} f_{\c \delta}
\end{eqnarray}
We note that, using the fact that $\hat{f}_{\a \b}$ is symmetric and gamma trace free, it can be written as
$\hat{f}_{\a \b} = \left(\c_{edcba}\right)_{\a \b} J^{abcde} $, which we will later use.

\subsection*{The relation between $\L$ and $\l$}

We now would like to make the transition to the standard fields $\l$ and $\hat{f}$. The conventional
spinor $\l$ is defined as the $\c$-trace of $\hat{f}_{\a b}$:
\begin{eqnarray}
\l^\a &=& \left(\c^c\right)^{\a \b} \hat{f}_{\b c}
\end{eqnarray}
To analyse $\hat{f}_{\a\b}$ up to order 5 in fields, we will need to evaluate $\l$ up to
order three in fields. Using the definition of the conventional spinor and some gamma matrix algebra we find
\begin{eqnarray}
\l^\b &=& \left(\c^b\right)^{\a \b}\hat{f}_{\a b} \\
&\stackrel{*}{=}& -10i \L^\b + \frac{14 i}{3} \L^\b \text{tr}\left(\cF\cF\right) - \frac{i}{3} \left( \c^{turs}\L\right)^\b
-\frac{1}{6} \partial_b \L\c^c\L \left( \c^b \c_c \L\right)^\b\\
&& + \frac{i}{12} \L\c_b{}^{rs} \L \left(\c^b\right)^{\a \b}D_\a \cF_{rs} \notag
\end{eqnarray}
Note that, up to this order, we may substitute $f_{ab}$ for $\cF_{ab}$. We solve for $\L$ in terms of $\l$ by recursion, 
dropping terms of order greater than three in fields. We also use the super Maxwell Bianchi identities to simplify the 
$D_\a \cF_{rs}$ term (the physical field strength, i.e. the one with the shifted potential, is identical to $f_{rs}$ at this order).
\begin{eqnarray}
\L &\stackrel{*}{=}& i \left[\l^\b + \frac{14}{30} \l^\b \text{tr}\left(\cF\cF\right) - \frac{1}{30} \left( \c^{turs}\l\right)^\b
+\frac{i}{60} \partial_b \l\c^c\l \left( \c^b  \c_c \l\right)^\b \right.\\
&& + \left. \frac{i}{60} \l \c_b{}^{rs} \l \c^{}_r \partial_s \l\right] \notag
\end{eqnarray}

As with the recursion to obtain $h$, here again one can derive the exact result by recursively substituting $\l$ 
for $\L$ since the recursion terminates due to the vanishing of the product of 16 spinors.

\subsection*{$\hat{f}_{\a \b}$ in conventional fields}

After having derived the necessary relations, we can now proceed as described earlier: plug the expressions for $\L$, $\cF$ and
$D_{\a} \L^\b$ into (\ref{foddodd}), the expression for $f_{\a\b}$. Then $f_{\a\b}$ will appear on the rhs and lhs of the equation.
Substitute the rhs for $f_{\a\b}$ on the rhs, and then impose the conventional constraint. The result may then be simplified. To do
so, one utilises the fact that $\c^{\left(3\right)}$ and $\c^{\left(7\right)}$ (a basis for the antisymmetric 16 by 16 matrices) 
are dual to each other. So any antisymmetric expression may be written in terms of $\c^3$ only. The result is:
\begin{eqnarray}
\hat{f}_{\a \b} \stackrel{*}{=} \left(\c^{abcde}\right)_{\a\b} &&\left( \frac{160}{3} \l\c_{edc}\l \hat{f}_{ba} \right.\\ \notag
&-& \frac{7544}{9} \l\c_{edc}\l \hat{f}_{ba} \text{tr}\left(\hat{f}\hat{f}\right)\\ \notag
&+& \frac{3508}{9} \l\c_{edc}\l \left(\hat{f}\hat{f}\hat{f}\right)_{ba}\\ \notag
&-& \frac{208}{9} \l\c_{ed}{}^{f}\l \left(\hat{f}\hat{f}\right)_{fc}\hat{f}_{ba}\\ \notag
&-& 4 \l\c_{e}{}^{gf}\l \hat{f}_{fg} \hat{f}_{dc} \hat{f}_{ba}\\ \notag
&+& 736 \l\c_{e}{}^{gf}\l \hat{f}_{fd} \hat{f}_{gc} \hat{f}_{ba}\\ \notag
&+& \frac{20 i}{3} \l\c_{ed}{}^{f}\l \hat{f}_{fc} \partial_b \l\c^a\l\\ \notag
&+& 56 i \l\c_{edc}\l \hat{f}_b{}^f \partial_{\left[\right.f} \l\c_{\left.a\right]}\l\\ \notag
&+& \frac{80 i}{9} \l\c_{edc}\l \partial_{\left[\right. b} \left[ \l\c_{\left.a\right]}{}^{gf}\l \hat{f}_{fg}\right]\\ \notag
&+& \left. \frac{i}{90} \l\c_f\c_{edcba}\c^{gh}\c^f\c^i{}_j\left(\partial_k\l\right) \l\c_i{}^{jk}\l \hat{f}_{hg}\right)
\end{eqnarray}
where $\stackrel{*}{=}$ indicates truncation on the rhs after order 5 in fields.

\section{Conclusion}
We have obtained the N=1, D=10 flat supersymmetric field strength constraint for abelian BI theory up to order five in fields, 
which corresponds to order six in an action. This theory with one linearly and one non-linearly realised supersymmetry is unique, as 
the only covariant constraint compatible with the Maxwell multiplet is the $\cF$-constraint. Only the order three term 
\begin{eqnarray}
\l\left(\c^3\right)\l f \label{order3}
\end{eqnarray}
had been known previously. In fact, at this order the constraint is fixed on dimensional grounds. The lowest order corrections to
the equations of motion and the associated Lagrangian terms were calculated in \cite{cnt1}. The equations of motion to next order
may be calculated from our result using the Bianchi identities. This calculation seems rather long and simplifying the resulting 
expressions is not an easy task.\\ 
It is tempting to try to generalise our approach to the non-abelian case. To that end, one might try to make the fields $\l$ and 
$\hat{f}_{ab}$ U(N) valued and use the techniques developed in \cite{cnt1} and \cite{cnt2}, where the lowest order 
correction was considered. That way one would of course not produce any commutator terms, so this approach would presumably be
incomplete.\\
Another possibility might be to write down a non-abelian version of the $\cF$-constraint. However, this would require the target 
space field $H$ to be group valued, the interpretation of which remains unclear. To circumvent this problem, one might then
try to consider splitting up the U(N) into a trace and a SU(N) part, where one might hope to use the approach 
described in this paper for the U(1) part. One would expect a deformation of the $\cF$-constraint in this case, however.\\
This splitting of the gauge group has also been considered in \cite{D0sorokin}, where a $\kappa$-invariant action for coincident
D0-branes was constructed.
\section*{Acknowledgments}

I thank Paul Howe and James Drummond for many useful discussions.

\bibliography{ThePaper07}

\end{document}